\documentclass[aps,prl,floats,twocolumn,showpacs]{revtex4}
\usepackage{epsfig}
\usepackage{amssymb}
\usepackage[dvips]{color}     
\usepackage[latin1]{inputenc} 
\usepackage{graphicx}

\newcommand{\nc}{\newcommand}
\nc{\tcb}{\textcolor{blue}}
\nc{\tcr}{\textcolor{red}}
\nc{\be}{\begin{equation}}
\nc{\ee}{\end{equation}}
\nc{\bea}{\begin{eqnarray}}
\nc{\eea}{\end{eqnarray}}
\nc{\bt}{\begin{tabular}}
\nc{\et}{\end{tabular}}
\nc{\ba}{\begin{array}}
\nc{\ea}{\end{array}}
\nc{\dy}{\displaystyle}
\nc{\pr}{{\rm I}}
\nc{\se}{{\rm II}}
\nc{\w}{{\rm w}}
\nc{\s}{{\rm s}}
\nc{\um}{{1\over 2}}
\nc{\Hc}{{{\cal H}_{\rm c}}}

\begin{document}

\title{Magnetic properties of two-phase superconductors}

\author{E. Di Grezia}
\email{digrezia@na.infn.it}
\affiliation{Universit\`a Statale di Bergamo, Facolt\`a di
Ingegneria, viale Marconi 5, 24044 Dalmine (BG), Italy
and Istituto Nazionale di Fisica Nucleare, Sezione di Milano, via
Celoria 16, I-20133 Milan, Italy}
\author{S. Esposito}
\email{salvatore.esposito@na.infn.it}%
\affiliation{Dipartimento di Scienze Fisiche, Universit\`{a} di
Napoli ``Federico II'' and Istituto Nazionale di Fisica Nucleare,
Sezione di Napoli, Complesso Universitario di Monte S. Angelo, via
Cinthia, I-80126 Napoli, Italy}
\author{G. Salesi}
\email{salesi@unibg.it} %
\affiliation{Universit\`a Statale di Bergamo, Facolt\`a di
Ingegneria, viale Marconi 5, 24044 Dalmine (BG), Italy;
and Istituto Nazionale di Fisica Nucleare, Sezione di Milano, via
Celoria 16, I-20133 Milan, Italy}

\begin{abstract}
We have recently proposed a theoretical model for superconductors
endowed with two distinct superconducting phases, described by two
scalar order parameters which condensate at different critical temperatures.
On analyzing the magnetic behavior of such systems, we have found some
observable differences with respect to the case of ordinary Ginzburg-Landau
superconductors. In particular, at low temperature the London penetration
length is strongly reduced and the Ginzburg-Landau parameter $\kappa$ becomes a
function of temperature. By contrast, in the temperature region between the two phase
transitions $\kappa$ is constant and the system is a type-I or a type-II
superconductor depending on the ratio between the critical temperatures.

\


\end{abstract}

\maketitle

\section{Two-phase superconductors}

In the Ginzburg-Landau (GL) theory \cite{GL, Annett, L9} the superconductivity is ruled by a complex order parameter, which can be interpreted as the wave function of the Cooper pairs in its center-of-mass frame. This order parameter undergoes a ``condensation'' when the temperature approaches a critical temperature. By extending the GL theory, we introduced in \cite{TwoTC, Term2TC} \textsl{two} order parameters, namely two scalar charged fields $\phi_\w$ and $\phi_\s$, to which correspond Cooper pairs where the electrons are bound by a weaker (\textit{w}) or stronger (\textit{s}) attractive force, respectively.
Let us stress that the introduction of additional degrees of freedom (two complex fields rather than one) \emph{does not involve new unknown physical constants} since both scalar fields are endowed with equal \emph{bare} masses and self-interaction coupling constants.

We started from the following Lagrangian (hereafter $\hbar=c=1$)
    \begin{eqnarray}
    {\cal L} =\left(D_{\mu }\phi_\w \right)^{\dagger }\left( D^{\mu }\phi_\w
    \right) + m^2\phi_\w^\dagger\phi_\w - \frac{\lambda}{4}(\phi_\w^\dagger\phi_\w)^2+
    \nonumber\\
    +
    \left(D_{\mu }\phi_\s \right)^{\dagger }\left( D^{\mu }\phi_\s
    \right) + m^2\phi_\s^\dagger\phi_\s - \frac{\lambda}{4}(\phi_\s^\dagger\phi_\s)^2 -
    \nonumber\\
    -\frac{1}{4}F_{\mu\nu}F^{\mu\nu}\,.\hfill
    \label{Ellex2}
    \end{eqnarray}
where \ $m^2>0$, \ $F_{\mu\nu}\equiv\partial_\mu A_\nu - \partial_\nu A_\mu$ is the electromagnetic field strength, \ and \ $D_{\mu }\equiv\partial_{\mu }+2ieA_{\mu }$ \ is the covariant derivative
($2e$ is the electric charge of a Cooper pair). At a (larger) critical temperature $T_1$ it occurs the condensation of the \emph{modulus} of the weak field $\phi_\w$; while at a (smaller) critical
temperature $T_2$ we have the condensation of the \emph{real part} (or \emph{imaginary part}) of $\phi_\s$. In \cite{TwoTC}
we showed that these different condensations of electrons in Cooper pairs do exhibit different \textit{effective} self-interactions. As a matter of fact, lowering the temperature of the system we meet a first spontaneous symmetry breaking (SSB) at a critical temperature $T_1$, and the medium becomes superconducting (hereafter we shall speak of ``phase-I''). By further lowering the temperature at $T=T_2$ the condensation involving the second order-parameter is energetically favored and a new (second-order) phase transition starts (``phase-II"). Below $T_2$ the system is more superconducting with respect to GL standard superconductors, since in addition to the ``weakly-coupled'' Cooper pairs, we should observe also the formation of ``strongly-coupled'' Cooper pairs.

After the condensations due to the U(1) SSB, the mean total free energy results as the sum of contributions from normal-conducting electrons, and from weakly-coupled and strongly-coupled Cooper pairs:
\begin{eqnarray}
F = F_{\rm n} & \ \mbox{\rm for} \ & T>T_1\,,
\label{Norm}
\nonumber\\
F = F_{\rm n} + a_\w(T){\eta_0}^2 + \frac{\lambda}{4}\,{\eta_0}^4 &
\ \mbox{\rm for} \ & T_2<T<T_1\,, \ \ \
\label{I}
\nonumber\\
F = F_{\rm n} + a_\w(T){\eta_0}^2 + \frac{\lambda}{4}\,{\eta_0}^4 +
\nonumber\\
+ a_\s(T){\chi_0}^2 + \frac{\lambda}{4}\,{\chi_0}^4 & \ \mbox{\rm for} \ & T<T_2\,,
\label{I+II}
\end{eqnarray}
where $\eta_0$ indicates the nonzero expectation value of $|\phi_\w|$; $\chi_0$ indicates the nonzero expectation value of ${\rm Re}\{\phi_\s\}$ (or Im$\{\phi_\s\}$), and \cite{Bailin, NXC}
\be
a_\w = - m^2 + \frac{\lambda + 4e^2}{16}\,T^2 \qquad a_\s = - m^2 + \frac{\lambda + 3e^2}{12}\,T^2
\ee
which vanish for
\be
T^2=T_1^2=\frac{16m^2}{\lambda+4e^2} \qquad\quad T^2=T_2^2=\frac{12m^2}{\lambda+3e^2}
\ee
respectively; notice that it is always \ \mbox{$1<(T_1/T_2)^2<4/3$.} \
In the above-quoted papers we found some deviations in basic thermodynamical quantities with respect to GL one-phase superconductors. In particular, in contrast to the usual case where only one jump in specific heat takes place at the normal-superconductor transition temperature, we predicted an additional discontinuity for $C_V$ when passing from a superconducting phase to the other one.

In the next section we shall study the behavior of the two-phase superconductors in the presence of an external electromagnetic field while varying intensity of magnetic field and temperature.

\section{Magnetic properties}

Let us start with the {\em Meissner effect}, namely the rapid decaying to zero of magnetic fields in
the bulk of a superconductor. The distance from the surface beyond which the magnetic field vanishes is known as the \textit{London Penetration depth}, and can be written in terms of the effective (after SSB and Higgs mechanism \cite{Naddeo, Tinkham, Higgs}) photon mass
$$
\delta = \frac{1}{m_A} = \frac{1}{8e^2|\phi_{\rm min}(T)|^2}
$$
where $\phi_{\rm min}(T)$ is the expectation value of the field in the minimum energy state
at a given temperature. Exploiting Lagrangian (\ref{Ellex2}), in the phase-I ($T_2<T<T_1$) we therefore have
\be
\delta_\pr = \frac{1}{8e^2|\eta_0(T)|^2}\,,
\ee
whilst in the phase-II ($T<T_2$) we now have two contributions to the photon mass
\be
\delta_\se = \frac{1}{8e^2(|\eta_0(T)|^2 + |\chi_0(T)|^2)}\,.
\ee
The expectation values $\eta_0(T)$ and $\chi_0(T)$ can be expressed \cite{Term2TC} as functions
of the critical temperatures
\be
\eta_0^2(T) = -\frac{2a_\w(T)}{\lambda} = \frac{T_2^2(T_1^2-T^2)}{24(T_1^2-T_2^2)}\,,
\ee
\be
\chi_0^2(T) = -\frac{2a_\s(T)}{\lambda} = \frac{T_1^2(T_2^2-T^2)}{24(T_1^2-T_2^2)}\,.
\ee
As a consequence we can write the London penetration lengths as follows:
\begin{equation}
\delta_\pr =  \left[\frac{e^2T_2^2T_1^2)}{3(T_1^2-T_2^2)}\left(1-\frac{T^2}{T_1^2}\right)\right]^{-\um}
\end{equation}
for the phase-I; and
\begin{equation}
\delta_\se = \left\{\frac{e^2T_2^2T_1^2)}{3(T_1^2-T_2^2)}\left[\left(1-\frac{T^2}{T_1^2}\right)
+\left(1-\frac{T^2}{T_2^2}\right)\right]\right\}^{-\um}
\end{equation}
for the phase-II.
Let us stress that, below the second critical temperature $T_2$, the penetration length of the magnetic field is smaller with respect to the GL one-phase superconductors.

The \textit{coherence length}
$$
\xi=  \frac{1}{m_\phi(T)}
$$
measures the distance over which the scalar field varies sensitively and is related to the mean binding
length of the electrons in a Cooper pair.
The coherence length can be expressed (via Higgs mechanism \cite{Naddeo, Higgs, Tinkham}) as a function of the temperature-dependent effective mass of the scalar field which results from a quantum field calculation including one-loop radiative correction: namely
\begin{equation}
m_{\phi_\w}(T) = \sqrt{- a_\w(T)}
\end{equation}
for the weak field ($T<T_1$); and
\be
m_{\phi_\s}(T) = \sqrt{- a_\s(T)}
\ee
for the strong field ($T<T_2$).
As expected, for $T\neq 0$,
$$
m_{\phi_\w}^2=m^2 - \frac{\lambda + 4e^2}{16}\,T^2 \,>\, m_{\phi_\s}^2 =m^2 - \frac{\lambda + 3e^2}{12}\,T^2
$$
since the (negative) binding energy between the electrons is larger for the strongly-coupled Cooper pairs.

Applying the above definition we get \textsl{two different coherence lengths} for the two fields
\begin{equation}
\xi_\w(T) = \frac{1}{m_{\phi_\w}(T)}\,,
\end{equation}
\begin{equation}
\xi_s(T) = \frac{1}{m_{\phi_\s}(T)}\,.
\end{equation}
At absolute zero the renormalized masses are equal to the bare mass $m$ and the two coherence lengths reduce to the common value:
\begin{equation}
 \xi_0 = \frac{1}{m} = 2\frac{\sqrt{4 T_2^2-3T_1^2}}{eT_1T_2}\,.
 \label{p0}
\end{equation}
Let us write explicitly the time dependence of the coherence lengths for the two types of Cooper pairs :
\begin{equation}
\xi_\w(T)=  \frac{1}{m_{\phi_\w}(T)}=\frac{\xi_0}{\dy\sqrt{1 -\left(\frac{T}{T_1}\right)^2}}
\qquad\quad T<T_1\,,
\label{p1}
\end{equation}
\begin{equation}
\xi_\s(T)=  \frac{1}{m_{\phi_\s}(T)}=\frac{\xi_0}{\dy\sqrt{1 -\left(\frac{ T}{T_2}\right)^2}}
\qquad\quad T<T_2\,.
\label{p1}
\end{equation}

\

\

\noindent Increasing the intensity of the magnetic field entering type-I superconductors, when
$H$ reaches a critical value $\Hc$, perfect diamagnetism and superconductivity are suddenly destroyed through a first-order phase transition. The critical magnetic field measures the ``condensation energy", given by the difference between the free energies of the normal and superconducting states
$$
F - F_{\rm n} = - \um\mu_0\Hc^2
$$
Exploiting the above equation we obtain the critical field in the phase-I, for $T_2<T<T_1$:
\begin{equation}
\Hc^\pr = \sqrt{\frac{2}{\mu_0\lambda}}\,|a_\w|
= \frac{eT_2^2(T_1^2-T^2)}{2\sqrt{6\mu_0(4T_2^2-3T_1^2)(T_1^2-T_2^2)}}\,,
\end{equation}
while in the phase-II, for $T< T_2$, we have
$$
\Hc^\se = \sqrt{\frac{2}{\mu_0\lambda}\left(a_\w^2+a_\s^2\right)} =
$$
\begin{equation}
= \Hc^\pr\sqrt{1+\frac{T_1^4(T_2^2-T^2)^2}{T_2^4(T_1^2-T^2)^2}}\,.
\end{equation}
Let us notice that there is no discontinuity at $T_2$: \ $\Hc^\pr(T_2)= \Hc^\se(T_2)$;
but, while for $T\lesssim T_1$ the critical field  decreases linearly, for $T\lesssim T_2$ we have instead a quadratic behavior
\begin{equation}
\Hc^\se \simeq \Hc^\pr(T_2)\left[ 1 + \frac{2T_1^4}{T_2^2(T_1^2-T^2)^2}(T_2-T)^2\right]\,.
\end{equation}

\

\

\noindent For type-II superconductors there exist two different critical fields, $\Hc_1$, the \textsl{lower critical field}, and $\Hc_2$ the \textsl{upper critical field}. We observe perfect diamagnetism only applying a field $H<\Hc_1$ whilst, when \mbox{$\Hc_1<H<\Hc_2$}, nonsuperconducting vortices can arise in the bulk of the medium. Abrikosov \cite{L9, Abrikosov} showed that a vortex consists of regions of circulating supercurrent around a small central normal-metal core: the
magnetic field is able to penetrate through the sample inside the vortex cores, and the circulating currents serve to screen out the magnetic field from the rest of the superconductor outside the vortex.

In the present model there are actually two different coherence lengths for two different types of Cooper pairs. Correspondingly, in the phase-II we shall have \textsl{different upper and lower critical fields} in ``domains" of the sample occupied by Cooper pairs of the same type, either weakly-coupled or strongly-coupled. As a consequence, in a given domain we can have (or not) Abrikosov vortices depending on the type of scalar field condensated in that domain. Such an inhomogeneity of the spatial distribution of the vortices in two-phase superconductors should result in a net, detectable difference with respect to GL superconductors: in principle, for $T<T_2$, a section of the material should show vortical and non-vortical sectors, in contrast to the homogeneous distribution of vortex cores in ordinary superconductors.

\

Starting from the London equation we can easily obtain the explicit expression of the lower critical field \cite{Annett, L9}
$$
\Hc_1 = \frac{\Phi_0}{4\pi\mu_0\delta^2}\ln{\frac{\delta}{\xi}}
$$
where $\Phi_0\equiv \dy\frac{\pi\hbar c}{e}$ is the so-called quantum magnetic flux unit. Thus, in the phase-I, 
the lower critical field writes:
\begin{equation}
\Hc_1^\pr = \frac{\Phi_0}{4\pi\mu_0\delta_\pr^2}\ln{\left(\frac{\delta_\pr}{\xi_\w}\right)} = h_1\left(1-\frac{T^2}{T_1^2}\right)
\end{equation}
where
\be
h_1\equiv\frac{\Phi_0}{4\pi\mu_0}\ln\left[\sqrt{\frac{6(T_1^2-T_2^2)}{4T_2^2-3T_1^2}}\frac{e^2T_1^2T_2^2}{3(T_1^2-T_2^2)}\right]\,.
\ee
As abovesaid, in the phase-II 
we have two distinct $\Hc_1^\se$:
$$
\Hc_{1_\w}^\se = \frac{\Phi_0}{4\pi\mu_0\delta_\se^2}\ln{\left(\frac{\delta_\se}{\xi_\w}\right)} =
h_1\left[\left(1-\frac{T^2}{T_1^2}\right)+ \left(1-\frac{T^2}{T_2^2}\right)\right] \times
$$
\be
\times \dy\left\{1-\frac{\ln\left[1 + \frac{T_1^2(T_2^2-T^2)}{T_2^2(T_1^2-T^2)}\right]}{\ln\left[\frac{6(T_1^2-T_2^2)}{4T_2^2-3T_1^2}\right]}\right\}
\ee
in the weak-field domains; and
$$
\Hc_{1_\s}^\se = \frac{\Phi_0}{4\pi\mu_0\delta_\se^2}\ln{\left(\frac{\delta_\se}{\xi_\s}\right)} =
h_1\left[\left(1-\frac{T^2}{T_1^2}\right)+ \left(1-\frac{T^2}{T_2^2}\right)\right] \times
$$
\be
\times \dy\left\{1-\frac{\ln\left[1 +
\frac{T_2^2(T_1^2-T^2)}{T_1^2(T_2^2-T^2)}\right]}{\ln\left[\frac{6(T_1^2-T_2^2)}{4T_2^2-3T_1^2}\right]}\right\}
\ee
in the strong-field domains.

\

Let us now pass to the upper critical field which can be expressed as follows \cite{Annett, L9}
$$
\Hc_2 = \frac{\Phi_0}{2\pi \mu_0}\frac{1}{\xi^2}\,.
$$
In the weak-field ($T<T_1$) and strong-field ($T<T_2$) domains the above equation writes, respectively:
\begin{equation}
\Hc_{2_{\rm w}} = \frac{\Phi_0}{2\pi\mu_0}\frac{1}{\xi_\w^2}=h_2\left(1-\frac{T^2}{T_1^2}\right)
\qquad\quad T<T_1\,,
\end{equation}
\be
\Hc_{2_{\rm s}} = \frac{\Phi_0}{2\pi\mu_0}\frac{1}{\xi_\s^2}=h_2\left(1-\frac{T^2}{T_2^2}\right)
\qquad\quad T<T_2\,,
\ee
\be
h_2 \equiv \frac{\Phi_0}{\pi\mu_0}\frac{e^2T_1^2T_2^2}{4T_2^2-3T_1^2}\,.
\ee
For $T\rightarrow 0$
\begin{equation}
\Hc_{2_{\rm w}} = \Hc_{2_{\rm s}} = h_2\,;
\end{equation}
for $T \to T_2$
\begin{equation}
\Hc_{2_{\rm w}} \neq \Hc_{2_{\rm s}} = 0.
\end{equation}

\begin{figure}[!h]
\centerline{\hbox{\psfig{figure=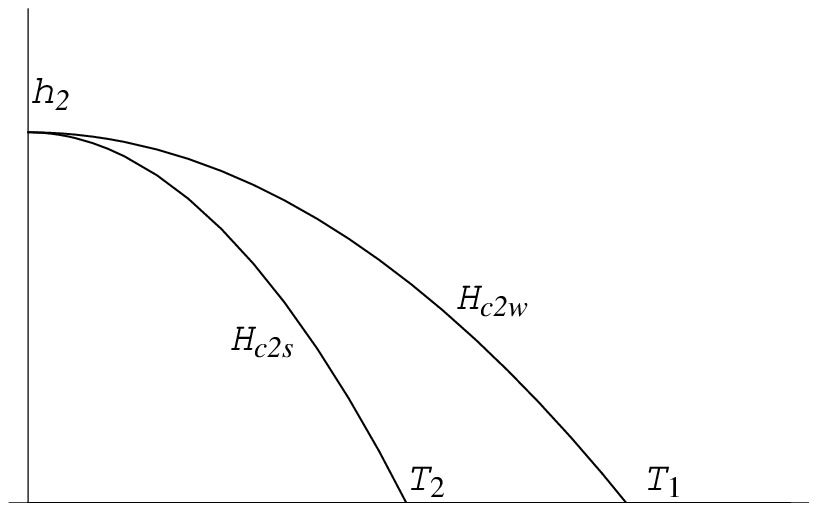,width=0.45\textwidth}}}
\caption{Upper critical fields vs. temperature for weak-field
and strong-field domains}
\label{fig1}
\end{figure}

The well-known dimensionless \textsl{Ginzburg-Landau parameter}
$$
\kappa \equiv \frac{\delta}{\xi}
$$
determines whether a medium is a type-I or type-II superconductor: for $\kappa<1/\sqrt 2$ it is a type-I superconductor; whilst for $\kappa>1/\sqrt 2$ it is a type-II superconductor.

In the phase-I we find:
\begin{equation}
\kappa_{\,\pr} = \sqrt{\frac{6(T_1^2-T_2^2)}{4T_2^2-3T_1^2}}
\end{equation}
which is independent of temperature in agree with the GL theory.  On imposing the condition $\kappa_{\,\pr}<1/\sqrt 2$ we infer that the material, for $T_2<T<T_1$, is a type-I superconductor if $\dy1\leq\left(\frac{T_1}{T_2}\right)^2\leq \frac{16}{15}$; and a type-II superconductor if $\dy\frac{16}{15}\leq\left(\frac{T_1}{T_2}\right)^2\leq \frac{4}{3} $.

In the phase-II we easily get:
\be
\kappa_{\se{\rm w}} = \left[1+\frac{T_1^2(T_2^2-T^2)}{T_2^2(T_1^2-T^2)}\right]^{-\um}\!\!\!\!\kappa_{\,\pr}
\ee
for the weak-field domains; and
\be
\kappa_{\se{\rm s}} = \left[1+\frac{T_2^2(T_1^2-T^2)}{T_1^2(T_2^2-T^2)}\right]^{-\um}\!\!\!\!\kappa_{\,\pr}
\ee
for the strong-field domains.

The dependence of $\kappa_\se$ on temperature could be observed as a net deviation from the  temperature-independent behavior of GL superconductors. Notice that $\kappa_\se<\kappa_\pr$:
therefore a type-II superconductor for temperatures in the region $T_2<T<T_1$ could become a type-I superconductor for $T<T_2$, if $\kappa$ decreases below  $1/\sqrt{2}$.
Quantities $\kappa_{\se{\rm w}}$ and $\kappa_{\se{\rm s}}$ turn out to be equal for $T\sim 0$:
\be
\kappa_{\se{\rm w}} = \kappa_{\se{\rm s}} \simeq \frac{\kappa_{\,\pr}}{\sqrt 2}\,.
\ee

\section{Conclusions}

In previous papers \cite{TwoTC} we introduced a model, in the framework of the GL theory, in order to describe physical systems experiencing two superconducting phases, whose critical temperatures $T_1, T_2$ differ at most of $15\%$. This model entails two different condensation regimes of two scalar fields with equal (bare) mass and self-interaction strength, describing Cooper pairs formed by differently interacting electrons, such different interaction arising from quantum loop corrections. In Ref.\,\cite{Term2TC} we have calculated the thermodynamical properties of such two-phase superconductors, the most peculiar being a second discontinuity in the specific heat; here, instead, we have focused on their magnetic properties. Below the second critical temperature, the penetration length of the magnetic field is smaller with respect to the usual one-phase superconductors (even of about $70\%$), depending on temperature. This is easily explained by the emergence of a second kind of Cooper pairs, in which electrons are more bonded than in the first kind of pairs, leading also to two distinct behaviors of the critical magnetic fields in the two superconducting phases. As a consequence, they exist two different coherence lengths for the electrons in the different Cooper pairs, resulting in peculiar superconductive properties. For instance, even if in the region between $T_1$ and $T_2$ the system is a type-II superconductor, depending on the ratio of the critical temperatures [$16/15\leq (T_1/T_2)^2\leq 4/3$], below $T_2$ it instead could behave as a type-I superconductor if $\kappa$ decreases sufficiently ($\kappa$ becoming $<1/\sqrt{2}$). Moreover, for $T<T_2$, the GL parameter $\kappa$ is not constant, and exhibits a characteristic dependence on the temperature, a result  strongly deviating from the predictions of the GL theory. Perhaps all these effects have not yet been observed in any material, due to the very small difference (no more than $15\%$) between the two critical temperatures, but this seems to be not a really difficult task for dedicated experiments, since those effects are very peculiar. Indications on this direction come from the known unusual properties of MgB$_2$. The predictions of our model cannot quantitatively apply to such superconductor system, mainly due to the large difference between the two critical temperatures measured, arising from fundamental physical effects different from what considered here \cite{2Gaps}. Nevertheless all the qualitative features of our model discussed above well apply to MgB$_2$, as it is readily realized by a comparison with the existing experimental literature (see, for example \cite{BuzeaYamashita} and Refs.\,therein). Several two-band theories try to explain the observed behavior of the specific heat of MgB$_2$ and the peculiar temperature-dependence of the upper and lower critical fields which result to be very similar to the ones predicted in our Ref.\,\cite{Term2TC} and in the present paper.
Maybe there is a correspondence between the \textsl{two} ``classical-macroscopic'' (since represent ``collective'' wave-functions for the condensate) order parameters in GL-like approaches as the present one, and the \textsl{two}  ``quantum-microscopic'' gaps in quasi-particle energy spectra predicted for MgB$_2$ by some BCS-like theories.\cite{2Gaps}
The experimental investigation on the novel kind of superconductors proposed here should then consist mainly in the careful search for materials which may exhibit the particular properties of the two kinds of electron pairs considered. The major merit of our model, with respect to the existing theories for (usual and) unusual superconductors is its full predictability, since all the basic physical properties are expressed in terms of the two measurable critical temperatures, rather than in terms of unknown quantities such as self-interaction coupling constants, quasi-particle effective masses, or mean distance between Cooper electrons.

\

{\bf Acknowledgements}

\noindent This work has been partially supported by I.N.F.N. and M.U.R.

\end{document}